\documentclass[english]{lni}
\usepackage{booktabs}
\usepackage{graphicx, cite}
\usepackage{amsmath, amssymb}

\begin{document}

\title{Semantic Enhancement of Lecture Material}

\author{Robin Nicolay\\
	University of Rostock\\
	Department of Computer Science\\
	Germany, Rostock 18055\\
	email: firstname.lastname@uni-rostock.de
}

\maketitle

\begin{abstract}
	Today's lectures are often talks following a straight line of slides. In many lectures the process of content teaching is not as efficient as it could be. Technologies, such as smart-phones and wireless communication, enable a new level of interaction between lecturer, content and audience. We describe how current lecture material can be semantically enhanced, to interactively assist the audience during and after a lecture.
\end{abstract}

\section{Introduction}
\label{sec:introduction}
In this paper, we describe how lecture material can be enhanced by adding semantic information. The contributions of this paper are:
\begin{itemize}
	\item Section \ref{sec:comparingpresentationwithbooks} discusses possible improvements in lecture material 
	\item Section \ref{sec:applications} introduces applications that allow an interaction between audience and lecture content 
	\item Section \ref{sec:enhancingsemantics} discusses possible semantic enhancements of lecture material 
	\item Section \ref{sec:topicmaps} introduces a model to build a semantic network over lecture material 
\end{itemize}

\section{Comparing presentations with books}
\label{sec:comparingpresentationwithbooks}
By comparing elements appearing in books such as encyclopedias or even Wikipedia \cite{Wikipedia.org.} with the process of content delivery during a lecture we identified a gap in the process of information exchange. We discuss differences to identify possible improvements of lecture slides and the process of giving a talk.

\subsection{A book - A structured network of information}
\label{sec:comparingbook}
Many educational books such as encyclopedias semantically highlight their content by using lemmas, paragraphs, text boxes and other elements. Books can easily be enriched by additional semantic information that highlight different classes of content or level of importance. Types of information such as examples, definitions and summaries help the reader to understand the purpose of information. An indication of importance of content helps target-oriented readers to focus on relevant information while additional information can be consumed by the reader when needed.

Beside semantic highlighting, semantic structures can help to improve the level of comprehension by guiding the reader to specific information if needed. Indexes and glossaries are structural elements that provide an overlaying network of information in addition to the straight ongoing path of a book. Readers can easily jump forward and backward or find annotated information. A structural overview provides an awareness of the objective and context of individual statements. Furthermore, it supports a less constrained path of information consumption. 

Semantic enrichment of information leading to contextual awareness assists in building an adequate mindset and find a right scheme \cite{Recker.1999}. This awareness helps readers to collect and adapt information correctly and sustainably.

\subsection{A lecture - A straight forward process}
\label{sec:comparinglecture}
Based on everyday observations of lecture contents, we noticed that lecture make much less use of semantic and structural highlights. Therefore, we discuss typical challenges faced by persons in lectures.

Keeping track of the agenda during a lecture can be a challenging task for the audience. The structural overview is essentially given by preliminary slides containing agendas and headings introducing new topics. Initial agenda slides and headlines introducing new topics mostly are not enough to provide a pervasive awareness about the current point in an argumentation path.

There are different use cases for slides in a presentation: Some illustrate facts; others show definitions or list summaries. Modern presentation tools such as PowerPoint \cite{Microsoft.2014} assist in marking these slides on base of layout and design. Nevertheless, listeners cannot always intuitively understand the slides' objectives. Information is rarely annotated or typed. Slides containing definitions and other references might be needed by the audience during later steps in a talk. Other slides provide assistance needed for understanding of subsequent facts. These requirements are typically not well supported by the straight forward nature of a talk.

Page numbers often are the only references to information and indication of progress. Page numbers are a linear system to keep track of references within a talk. Nevertheless, they do not provide easy-to-use linking or networking of topics, areas, or expressions within a talk.

\section{Applications to improve lecture learning}
\label{sec:applications}
The goal is to enhance current lecture slides to diminish the gap between semantically networked information in sources such as books and the straight forward process of attending a lecture.

Currently, a typical slide show consists of a linear set of slides. These slides are often used as additional visual channel which is extending the lecturer's talk with explaining illustrations or headwords. Slides ideally follow the lecturer as he speaks and are shown straight forward at the visual communication channel to the audience. Some approaches aim for supporting the dialog between lecturer and audience by providing a third, independent information channel (see \cite{Vetterick.2013} and \cite{TopHatMonocleInc..2014}). Our work focuses on optimization of the communication between a presentation's content and the audience.

There are two applications under development. Both applications will be published at a later point but are shortly mentioned here.

\begin{figure}
	\centering
	\fbox{
		\includegraphics[width=0.97\textwidth]{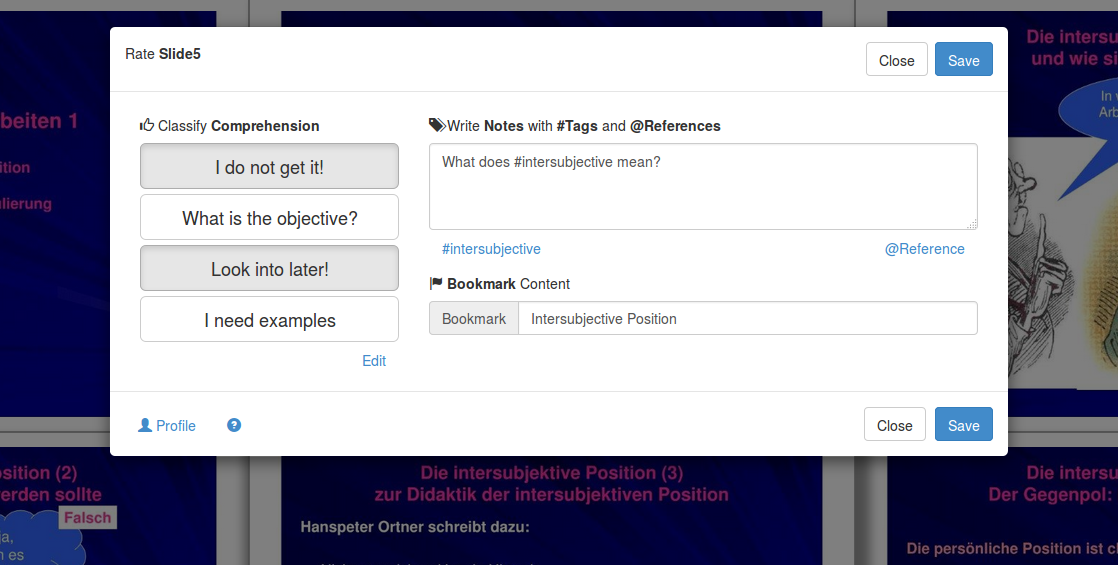}
	}
	\caption[LectureRating screenshot]{Current slides in a lecture are delivered to the audience via web client. Participants of the audience can classify comprehension, attach notes, tag content, link references or bookmark content}
	\label{fig:lecturerating}
\end{figure}

The first application that is developed in this context is a content rating system for mobile devices. With this system, ratable lecture content is delivered to the participants' mobile phones during the talk. The system collects the listeners' opinion. For every ratable piece of information, the user can use buttons to easily classify their grade of comprehension. Additionally he can add notes with the ability to tag topics and reference content. User assigned bookmarks help to track the lecture path and to provide an overview of lecture material. 

Ratings during the lecture help the individual listener to keep track of content that has to be post-processed after a lecture. Individuals of the audience sharing the same requirements find each other based on used comprehension classes and tags in notes. The use of tags enables an export of slides and issues to discussion groups after a lecture. Additionally, an aggregation of these information assist presenters of a lecture to find parts in his lecture that are difficult to understand and open overall questions that need further attention.

The first application is used to deduce the above mentioned annotations, classifications and references from observation of lecture participants' behavior. 
The second application assists the individual requirements with the help of lecture material.

The second application developed in this context can assist a lecture participants by supporting individual needs for assistance with a slide or paragraph during a talk. Supporting individual requirements in large audiences can be a very challenging task for a lecturer. Individual requirements can be caused e.g. by different previous knowledge and education in a heterogeneous audience. 
Our approach uses mobile devices to provide suitable auxiliary information if needed. The system provides suitable information such as set of references to other slides or passages being related with preliminary knowledge or classified as definitions or examples and part of the same subject the slide is a part of.

To facilitate individual support based on lecture material, lecture material, such as linear slide shows, needs to be semantically enhanced. The following sections describe how we enhance lecture content to support a autonomous assistance.

\section{Enhancing semantics in lecture materials}
\label{sec:enhancingsemantics}
By reviewing typical challenges faced by persons in a lecture we developed a concept to improve the semantic highlights of information and the relationship between single slide show elements. First, we provide a short overview on possible semantic relations, followed by an approach to transfer lecture material to a semantic network based representation.

Coming from how slide shows are currently used, we identified some straight forward semantic highlights.

\paragraph{Annotate information classes and objectives} As described in \ref{sec:comparingpresentationwithbooks} content can be enhanced by classifying the type of information. Classifications, such as announcements of new topics, definitions, summaries or even overviews or structural information help to understand the current objective of information in the flow of argumentation. Content classes defining the objectives of information furthermore assist in finding suitable types of information for a specific requirement during and after a lecture.

\paragraph{Highlight anchors and topics} Like chapters in a book, topics of a talk need an overall reference. These labels (we call them anchor) identify a context we currently talk about and help forecasting and identifying possible objectives. As stated in Schema Theory \cite{Recker.1999}, to learn and understand new information, it is beneficial to attach it to existing information. Starting a talk without providing any anchors or checkpoint will result in a lack of comprehension as well as a missing availability of relevant terminology to the current context. Anchors provide an overview of a talk's structure. Attached to current slides, they increase the awareness for the current context and necessary mindset.

Beside semantic highlights of information, we identified relations between pieces of information. These relations aim to build an overall network of semantic dependencies.

\paragraph{Temporal continuity} Topics of a lecture being close to each other in a presentation are not necessarily related. However, the temporal proximity may indicate some not closer defined relation in the argumentation path of the talk. Temporal continuity of topics can indicate some context, e.g. paths towards a topic.

\paragraph{Preliminary knowledge} During a lecture, content will be provided in series of information. Comprehension of, e.g. deductive information may need previous knowledge. A presenter structures content to provide necessary previous knowledge before presenting information. Therefore, knowledge needed for comprehension needs to be related with their subsequent information.

\paragraph{Approaching paths} There can be various paths to approach a topic. The way to approach a topic not only depends on the beforehand presented previous knowledge, but also on elements such as definitions or examples. These elements bridge the gap between the topic fact and a listener’s previous knowledge and context. A lecturer may have multiple definitions, examples and argumentative paths to communicate the very same issue to different listeners. Requirements for these elements may differ regarding the individual preferences and experiences of persons in an audience. Available, elements covering different perspectives should be related to a topic even if they are not part of the current talk.

\paragraph{Direct references} Referencing another topic might be a good way to attach relations on any aspect of two topics. Nevertheless, even two topics not being related to each other might share equality in some elements. These direct relations between facts can serve for sustainable comprehension in ways such as being a memory hook or a reference to similar knowledge.

\paragraph{Relating topics in scopes} We define a scope as a set of related topics and slides. Knowledge of a scope of a topic provides a good overview and defines the perspective a topic is used in. This perspective may influence the set of available approaching paths. Nevertheless, building a scope requires knowledge about the context of a topic and the topic itself. Therefore, awareness about scopes might help to identify correct schemes \cite{Recker.1999} and increase the level comprehension. On the other hand the ability to define a scope might provide a good indication for the level of comprehension.

\paragraph{Feedback schemes} A lecturer giving a talk expresses a mindset and context to the audience. Adapting the mindset of a lecturer is an essential part of delivering knowledge between a presenter and the audience \cite{Laurillard.1999}. It is essential that content is adapted in the context it should be. From the perspective of the audience information might be related in another context or scheme. Interpreting information on different contexts is driven through perspective dependent previous relations in the mindset of the audience. The aggregation of feedback of assumed context or scheme used by audience might be a good indicator for an inadequate presented mindset.

\section{Using Topic Maps to enhance semantics in lecture material}
\label{sec:topicmaps}
We use the following model to abstract a semantic net with topics and slides as nodes and functional classified vertices from the otherwise linear set of slides. The concept of Topic Maps is used to utilize the semantic improvements described in \ref{sec:enhancingsemantics}. We aim to increase the highlighting of types, the structural as well as contextual relationships between information. A lot of elements listed in \ref{sec:enhancingsemantics} can directly be inferred in a Topic Map representation.

Topic Maps support a global knowledge interchange as described in \cite[p.47]{Hunting.2003} and are described in \cite{Wandora.}. They consist of the following elements:

\begin{itemize}
\item Topics or subjects are entities you can talk about
\item Topics can have characteristics or types
\item Topics need to be identifiable (subject identifier e.g. by a URI)

\item Associations are semantic relations between topics
\item Associations can have types and member roles

\item Occurrences are anything from the real world that is related to a topic
\item Occurrences are slides, documents, media files et cetera
\item Occurrences can be typed

\item Scopes and facets define the context in which topics, associations ,and occurrences are used
\end{itemize}

\begin{figure}
	\centering
	\fbox{
		\includegraphics[width=0.97\textwidth]{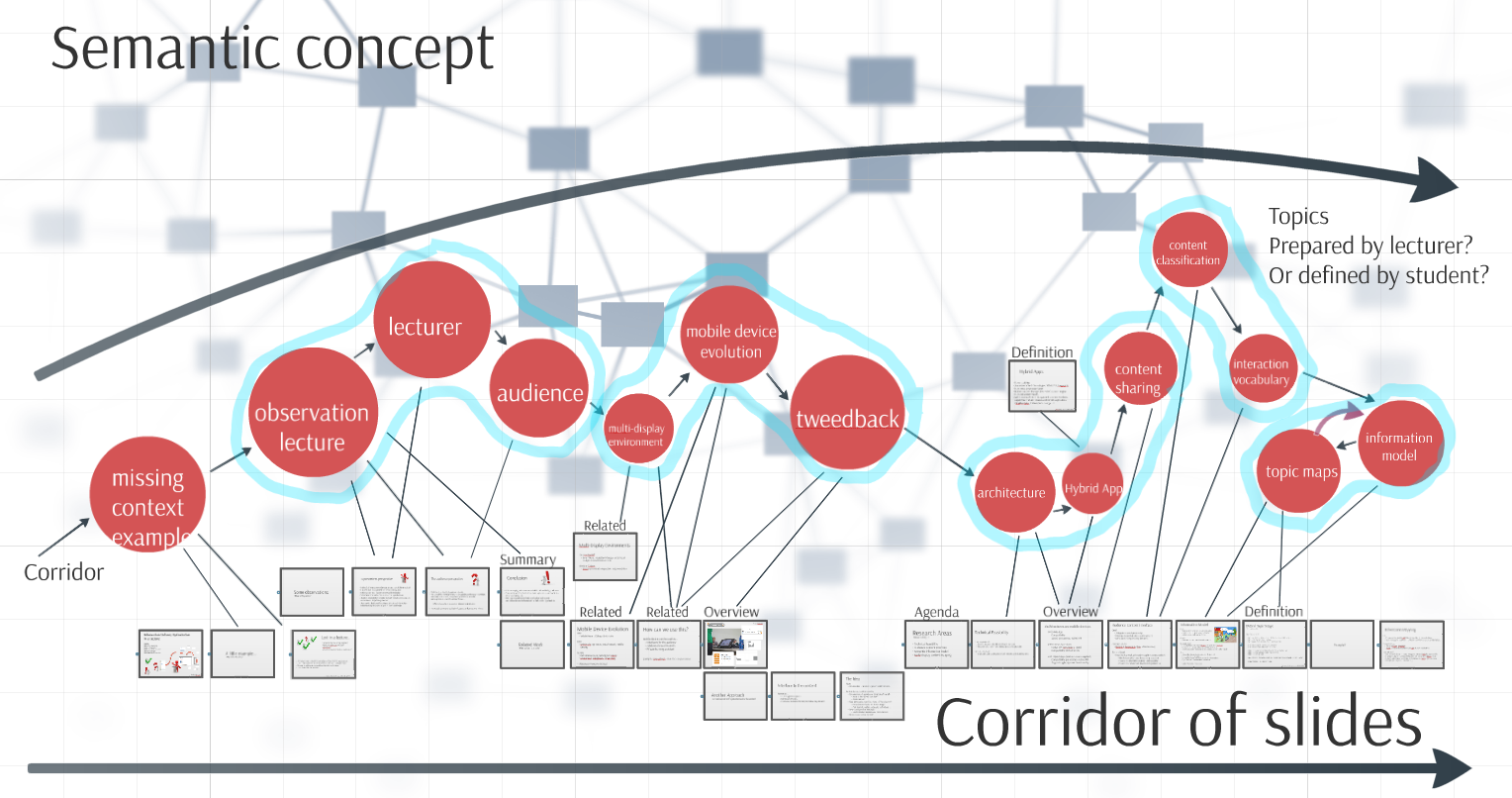}
	}
	\caption[Lecture material Topic Map]{As described in \cite{Wandora.} Topic Maps consist of topics and occurrences. Topics are represented as red dots and occurrences as grey boxes.}
	\label{fig:topicmap}
		
\end{figure}

A set of slides used in presentations often follow a straight argumentation path. We named it a “corridor of slides”. This corridor not only represents the path through a talk. Furthermore it reflects the mindset of the presenter. Information sources, such as Wikipedia allow a free digging into a set of information; however, a lecture is closely organized along a narrow route. The presenter of a lecture decides which information is relevant and therefore is part of the lecture. Our approach aims to preserve this path. As stated in \cite{Hembrooke.2003} providing assistance by connecting extensive knowledge management systems leads to distraction and less attention during a lecture.

\paragraph{Topics} Similar to the process of classifying scientific papers we suggest defining a set of keywords or tags. A keyword or tag can be used to describe a topic or name the area of information. We use keywords and tags to sum up the content or topic of a subset of slides. They can be seen as anchor or mindset headline (as described in \ref{sec:enhancingsemantics}) holding a set of slides together.

\paragraph{Associations} As shown in figure \ref{fig:topicmap} we use association between topics to draft the argumentation path of the lecture through a set of slides. The temporal continuity through a set of topics or anchors builds a path of related headlines denoting the agenda and structure of a lecture.

Furthermore, association can be used to describe semantic relations between topics such as preliminary knowledge. These associations describe an additional relationship between anchors and can be used to reference previous knowledge topics necessary to understand subsequent topics.

\paragraph{Occurrences} The occurrences of a topic are represented through a subset of related slides. Many slides can be annotated with their type (as described in \ref{sec:enhancingsemantics}) such as to introduce a new topic, provide definitions, examples, summaries, agendas or even overview information.

\paragraph{Scopes} Topics sharing the same set of occurrences build a scope marked in blue at figure \ref{fig:topicmap}. This scope can be seen as context of a topic. A topic being part of a scope is part of a context. The context defines the perspective the topic currently is placed in. One topic might be used in several contexts, argumentation paths and can be seen from several perspectives. 

Inferring presentation slides in topic maps enables the sharing of different contexts. One advantage of topic maps is the merge of different topics having the same identifier (keyword or tag) and their contexts. Including another set of lecture slides into this map, that share some topics, enables a global linking of topics and information. Furthermore, it enables a merging of argumentation paths, contexts, perspectives, and dependencies between topics from different lectures.

\section{Collecting semantic information for lecture material}
\label{sec:collectingsemantics}

\paragraph{Prepared Collection} As first step to collect semantic information for lecture material we develop an interface for the presenter. This interface enables the lecturer to import current sets of slides. Slides can be annotated with classes such as definition, summary, example to define their type and intention. Subsets of slides can be annotated with keywords and tags containing information such as headlines or anchors. These Keywords build the initial set of topics (nodes) that connect slides (occurrences) to a higher abstract layer. Additional slides that are not part of the lecture but share the same topics attach themselves to be available in the lecture. Additional semantic associations can be drawn to relate topics with semantic dependencies such as preliminary knowledge. The system automatically identifies contexts between different topics that share the same occurrence. The order of slides used in a talk define temporal continuity associations between topics.

\paragraph{Crowd Souring on Topics} The application of a lecture content interface mentioned in \ref{sec:applications} provides the ability to act with delivered content in a lecture. Content can be classified by comprehension and annotated with notes, tags, references, and bookmarks.

Classes help to sort information into a fixed set of equivalence classes. These classes for example can be used to classify content into a fixes set of comprehension levels. A fixed set of classes builds a manageable base for automatic consensual evaluations that help presenters to identify consensual lacks of comprehension within the talk. Additionally classes can be used to relate slides on behalf of their comprehensibility.

Tags and references can be used in versatile ways. They provide the ability to define anchors (topics) for later discussions based on rated content (occurrences). Tags and references will be used to extend the semantic network. In difference to fixed classification classes tags and references support to form individual overlying relations between information. Including the content of written notes, we build a crowd sourced set of new discussion topic nodes related to content of lecture slides. Freely build relations and topic nodes can also be used as an indication of the consensual context assumed by the audience.

Additionally bookmarks can be delivered by the presenter or be defined by the audience. Bookmarks provide an awareness of the current position in a talk and enable a fast referencing and retrieval of past content. Furthermore they provide the ability to define checkpoints in a lecture.

\section{Benefits and outlook}
\label{sec:benefitsandoutlook}
Introduced applications and concepts aim to improve lecture learning in order to:

\begin{itemize}
\item Improve the awareness for the current talk’s argumentation structure
\item Track the current position and perspective in current argumentation flow
\item Highlight the objective of currently presented content
\item Support different contextual backgrounds of the individuals in an audience and provide different perspectives of argumentative paths to the individual participant of a lecture
\item Help the students to define and communicate the correct mindset by an permanent awareness of context and topic
\item Support easy-to-use references between different parts of a lecture to retrieve information when needed
\item Help to relate lecture material to discussion topics in order to support discussions based on lecture content
\item Provide an overview of comprehension for participants based on rated lecture material
\item Provide an overview of overall comprehension in the audience for the presenter. Support to check the correlation of the mindset between lecturer and audience.
\end{itemize}

Our goal is to support a flexible context and mindset network. This network will merge different contexts of audience individuals. At a later point we use this information to match the previous defined mindset of a lecturer to the consensual extracted mindset of the audience. Information of correlation helps to dynamically improve lecture materials, argumentation paths and refine the corridor of argumentation in a lecture.

\section*{Acknowledgements}
\noindent This work is supported by the German Research Founda-
tion (DFG) as part of the graduate school MuSAMA (grant
no. GRK 1424/1).
Thanks to Andreas Daehn who helped to express my scatterbrained ideas.

\bibliographystyle{lni}
\bibliography{./lit}

\end{document}